# Queueing for Civility: User Perspectives on Regulating Emotions in Online Conversations

AKRITI VERMA*, Deakin University, Australia
SHAMA ISLAM, Deakin University, Australia
VALEH MOGHADDAM, Deakin University, Australia
ADNAN ANWAR, Deakin University, Australia

Online conversations are often interrupted by trolling, which causes emotional distress and conflict among users. Previous research has focused on moderating harmful content after it has been posted, but ways to manage emotions in real-time remain unexplored. This study suggests a comment queuing mechanism that delays comment publishing, encourages self-reflection, and reduces the impact of impulsive and toxic comments. To assess the efficacy of this approach, a mixed-method research design is used. An analysis of 15,000 user interactions on Reddit showed that this approach could reduce the spread of hate speech and anger by up to 15%, with only 4% of comments being delayed for about 47 seconds on average. We also surveyed users for feedback on the mechanism. The results showed that 93. 3% of the participants thought that the queuing mechanism could help calm the discussions and showed interest in seeing it used on social media platforms. Furthermore, 83% believed it would reduce impulsive comments and balance the emotional tone in conversations. We found a strong link between users' typical emotional states while using social media and their perceptions of the delay, with calm users finding the mechanism helpful and frustrated users anticipating frustration.

## 1 Introduction

Online conversations play a key role in shaping public discourse, encouraging discussions, and promoting social media engagement [Parameswaran and Whinston 2007], [Wadley et al. 2020]. However, they are also susceptible to toxicity, often fuelled by impulsive reactions, trolling, and emotionally charged interactions [Smith et al. 2022]. Research indicates that heightened emotional content, particularly anger and outrage, generates greater engagement than neutral content, facilitating the rapid spread of inflammatory discourse [Bodaghi and Oliveira 2022], [Brady et al. 2017]. This trend is further aggravated by algorithmic amplification on social media platforms, which often intensifies conflicts and hinders constructive dialogue [Solovev and Pröllochs 2022]. While current moderation techniques, such as content removal and user bans, aim to suppress

Authors' Contact Information: Akriti Verma, vermaakr@deakin.edu.au, Deakin University, Australia; Shama Islam, Deakin University, Australia, shama.i@deakin.edu.au; Valeh Moghaddam, Deakin University, Australia, valeh.moghaddam@deakin.edu.au; Adnan Anwar, Deakin University, Australia, adnan.anwar@deakin.edu.au.

toxic content after it has been posted, they fail to address the underlying issue i.e. the unregulated emotional responses that drive impulsive and toxic interactions [Chandrasekharan et al. 2022], [Trujillo and Cresci 2022], [Gongane et al. 2022].

Self-reflection is widely acknowledged as an essential process in emotion regulation, enabling individuals to evaluate their emotional states and reactions before taking action [Herwig et al. 2010]. In offline contexts, self-reflection provides individuals with the opportunity to assess how their emotions impact social interactions, leading to more thoughtful responses. However, in digital environments, where interactions occur rapidly and impulsively, users often lack the opportunity to pause and regulate their emotional responses before engaging in discussions [Kiskola et al. 2021]. This gap highlights the need for real-time intervention strategies that facilitate emotion regulation before a comment is posted, rather than relying exclusively on reactive moderation [Maarouf et al. 2022], [Goel et al. 2016].

To address this challenge, we propose a comment queuing mechanism that delays the posting of potentially harmful comments, allowing users time to reconsider their responses while allowing the conversation to evolve. This mechanism acts as an implicit digital emotion regulation (DER) strategy, creating a structured pause that promotes self-reflection and reduces the immediate escalation of negativity. By introducing a brief delay before a potentially disruptive comment is published, this mechanism disrupts the cycle of impulsive engagement that contributes to online toxicity.

We evaluated the effectiveness of a queuing mechanism using a mixed-method approach. Initially, we analysed 15,000 user interactions on Reddit to assess the impact of the queuing framework on the spread of emotional contagion in online discussions. Our findings indicate that this mechanism reduces the propagation of hate speech and anger by up to 15%, with only 4% of comments being temporarily held for an average of 47 seconds before publication.

To complement our analysis, we conducted a user survey to gather insights regarding perceptions of the queuing mechanism. The survey results show strong support for this approach, with 93.3% of participants believing it would help calm online discussions and mitigate both intentional and unintentional trolling. Additionally, 83% of respondents felt that the delay would curb impulsive comments and assist in balancing the emotional tone of conversations. However, perceptions of the queuing mechanism varied according to users' typical emotional states when engaging with social media. Those who reported feeling calm online viewed the mechanism positively, while those who frequently experienced frustration were more inclined to find the delay aggravating.

This study highlights the potential of self-reflection-based digital emotion regulation in transforming online interactions by integrating computational analysis with user feedback. Instead of relying exclusively on punitive moderation strategies, the queuing mechanism presents a proactive, user-centred approach that fosters healthier, more balanced conversations online. Therefore, this work makes the following research contributions:

- Comment Queuing for Digital Emotion Regulation: Proposes a comment queuing mechanism that introduces a delay for potentially toxic comments, fostering self-reflection and proactive emotion regulation in online discussions while reducing the likelihood of impulsive or hostile responses.
- Mixed-Method Evaluation of Online Moderation: Combines computational analysis of 15,000 Reddit interactions with user feedback, showing that delays can help reduce toxicity by up to 15% while maintaining conversational flow. User survey confirmed that 93% of respondents supported implementing the mechanism to calm conversations and prevent impulsive outbursts.

- Advancing Emotion-Aware Moderation: Demonstrates how dynamic toxicity thresholds and temporal interventions create a more balanced and adaptive approach to online content moderation.

The rest of this paper is structured as follows: It begins with a review of relevant literature (Section 2, Literature Review), followed by a description of the proposed framework (Section 3, Methodology). Then, the results of the mixed method evaluation are presented (Section 4, Results and Discussion) followed with the limitations (Section 5, Limitations) of this work before summarising the conclusion (Section 6, Conclusion and Future Work).

## 2 Literature Review

The challenge of addressing online toxicity and promoting constructive digital interactions has been widely researched. This section focuses on three essential areas: the nature and impact of trolling in online discussions, current strategies for emotional regulation in digital communication, and the potential of self-reflection as a tool for moderating emotions in online environments.

### 2.1 Trolling in Online Conversations

Trolling, defined as the intentional disruption of discussions through provocative or offensive messages, has been extensively studied within Human-Computer Interaction (HCI) research. Various studies highlight the motivations behind trolling, which may range from seeking amusement and attention to intentionally inflicting harm or steering discussions in a specific direction [Buckels et al. 2014]. The consequences of trolling include heightened emotional arousal, interpersonal conflicts, and the deterioration of productive discourse within online communities [Kumar et al. 2017], [Cheng et al. 2017]. Furthermore, trolling can escalate into more serious forms of online harassment and cyberbullying, resulting in significant emotional and psychological distress for those affected [Jane 2020].

Although there exists a substantial body of research examining the nature and consequences of trolling, much of the focus has been on post hoc moderation strategies, such as automated filtering systems and content removal policies [Yin et al. 2009], [Chandrasekharan et al. 2017], [Chandrasekharan et al. 2022]. While these reactive approaches contribute to maintaining safer online environments, they fail to address the root causes of trolling and do not prevent the emotional harm that accompanies such interactions.

### 2.2 Emotion Regulation in Digital Communication

Emotion regulation is the ability to manage and modify emotional responses. It is vital to foster constructive social interactions both offline and online. In digital environments, the absence of non-verbal cues can complicate emotional interpretation, increasing the risk of miscommunication or emotional escalation [Derks et al. 2008]. Research in this area has investigated various strategies for managing emotions in digital settings, including cognitive reappraisal, suppression, and the use of emojis or emoticons to clarify intent [Holtzman et al. 2017].

The impact of digital platforms on emotion regulation has been extensively examined. For example, social media platforms often amplify emotionally charged content due to algorithmic prioritisation, which can intensify emotional responses and contribute to a toxic environment [Roberts 2016], [Kramer et al. 2014]. Conversely, some platforms have integrated tools designed to assist with emotion regulation, such as content warnings, mute and block features, and reminders to take breaks [Schoenebeck 2014].

However, there is a gap in the literature regarding real-time emotion regulation strategies during digital interactions. Most existing solutions either intervene after an incident has occurred or depend on users actively choosing to manage their emotional responses (e.g., by muting a conversation) [Holtzman et al. 2017], [Kramer et al. 2014], [Schoenebeck 2014]. Limited research has focused on how platforms can be designed to encourage emotional self-regulation in real time before potentially harmful interactions arise.

### 2.3 Self-Reflection as a Tool for Emotion Regulation

Self-reflection is the process of examining and understanding one's emotions and behaviours. It has been extensively recognised in psychological research as an effective strategy for emotional regulation [Herwig et al. 2010]. Research indicates that self-reflection can enhance emotional awareness, reduce impulsivity, and facilitate thoughtful decision-making [Gross 2002], [Upadhyaya 2020]. Additionally, it has been associated with improved conflict resolution, with evidence suggesting that encouraging individuals to reflect on their emotions can reduce aggression and foster better interpersonal communication [Kross and Ayduk 2008], [Ayduk and Kross 2010].

Despite its numerous advantages, self-reflection remains underused as a digital intervention in real time [Kiskola et al. 2021], [Kiskola et al. 2022]. Although some digital tools, such as mindfulness apps, incorporate reflective practices, their application in moderating online discussions remains unexplored [Howells et al. 2016], [Ruckenstein and Turunen 2020], [Torre and Lieberman 2018]. Most interventions that integrate self-reflection are designed for post-interaction contemplation rather than immediate, in-the-moment emotional regulation.

This study aims to bridge this gap by introducing a comment queuing mechanism that promotes self-reflection in digital conversations. By incorporating a comment queuing system that delays the posting of potentially harmful content, we aim to encourage users to pause and consider both their emotional state and the potential consequences of their words. This method not only addresses the problem of online trolling but also promotes more thoughtful and intentional digital interactions.

### 2.4 Gaps in the literature

The key gaps that exist in the literature on self-reflection as a tool for emotional regulation:

- Lack of real-time emotion regulation strategies: Existing methods for tackling negative online behaviour, such as trolling, primarily rely on after-the-fact moderation, addressing issues only once they have already occurred. There is a gap in research on proactive strategies that can help manage emotional responses in real-time during online interactions.
- Limited integration of self-reflection in digital platforms: While self-reflection is a well-established emotion regulation strategy within psychology, its application in digital communication platforms remains unexplored. Further investigation is needed to determine how self-reflection can be integrated into online discussions to assist users in managing their emotions.
- Failure to address the root causes of trolling: Current moderation systems are primarily designed to detect and remove toxic content, overlooking the underlying emotional and behavioural patterns that contribute to trolling. There is a need for interventions that not only mitigate toxic content but also promote more thoughtful and constructive engagement among users.

## 3 Methodology

In this section, we present a strategy for implementing a comment queuing system aimed at promoting self-reflection and moderating emotional reactions in online discussions.

## 3.1 Core Methodological Insights

This paper presents a novel approach to promoting civility in online discussions by introducing a comment queuing mechanism. We propose intentionally pausing the act of posting to allow users time to reconsider or revise their thoughts. This design is informed by behavioural science, which suggests that brief pauses can help disrupt impulsive and emotionally charged reactions [Suler 2004], [Baumeister et al. 2007], [Gross 1998]. Our proposed framework allows the queuing of one or more comments while retaining control over when to release them. The queuing mechanism is designed for emotionally intense or conflict-prone conversations. By buffering real-time interaction, it helps reduce the escalation of hostile exchanges and provides users an opportunity to cool down before responding. In accordance with the framework established by [Slovak et al. 2023] for classifying technology-based emotion regulation interventions, the proposed queuing mechanism in our study can be positioned within its three core dimensions:

- **Theoretical Component**: This queue-based approach is designed to foster self-reflection prior to response, specifically in line with Gross's Process Model of Emotion Regulation [Gross 2008]. It encompasses response modulation and attentional deployment, prompting users to reconsider or modify their comments if flagged for exceeding toxicity thresholds.
- **Strategic Component**: The queuing mechanism employs an experiential learning strategy by introducing an on-spot intervention during the comment-posting process. It offers real-time feedback regarding the emotional tone of the conversation, by dynamically adjusting the queuing thresholds based on the conversation's emotion levels.
- **Practical Component**: In practical terms, this design utilises both implicit feedback mechanisms (queue time acting as a moderating factor) and explicit prompts (requests for comment revision) to encourage self-reflection and reduce the escalation of negative emotions.

## 3.2 Data Collection

To create a dataset for evaluating our proposed queuing mechanism, we gathered textual conversation data from Reddit, a popular discussion platform where users actively participate in debates, share opinions, and exchange experiences [Manikonda et al. 2018]. Given the focus of our study, we concentrated on discussions within political subreddits, as political conversations tend to be particularly engaging and often emotionally charged. The selected subreddits included r/politics, r/worldpolitics, r/politicaldiscussions, and r/politicaldebates, as these subreddit are known for eliciting strong emotional responses.

To reflect a diverse range of conversational structures and levels of engagement, we categorised discussions by their size, selecting threads that contained at least 2000, 3000, 5000, and 7000 comments. This classification enabled us to assess how the queuing mechanism performs in conversations with varying degrees of interaction. Data collection was conducted from August 2023 to August 2024, during which we retrieved 65 conversation threads. We utilised the Reddit API (PRAW) to extract post content, metadata, timestamps, and comment structures. To maintain the contextual relationships between comments, we preserved the parent-child relationships within the discussion threads, allowing for a structured graph representation for further analysis.

## 3.3 Data Preprocessing

After extracting the raw conversation data, we undertook a series of preprocessing steps to prepare and structure the dataset for analysis. Initially, we filtered out non-English discussions to ensure linguistic consistency and focused on posts containing textual content and emoticons only. Each conversation was organised in a structured format using CSV files, with individual files designated for each user. To clean the text data, we eliminated extraneous characters, hyperlinks, special symbols, and redundant spaces to enhance textual clarity. The textual content was segmented into smaller, meaningful units (tokens) to facilitate conversation analysis. Acknowledging the significant contribution of emoticons to the emotional tone of messages, we employed the Emojinal library [Barry et al. 2021] alongside Gensim to convert emoticons into numerical vector representations [Bird et al. 2009], [Verma et al. 2023].

For this study, we used a dataset containing 15,000 instances of user posts and interactions from Reddit. The dataset encompasses the following main components:

- Original Posts: These are the foundational nodes in conversation graphs.
- Replies and Comments: These become nodes that will be examined for their emotional content.
- Timestamps: We leverage timestamps to monitor the timing of each comment's submission.

For emotion classification, we used the NRC Word-Emotion Association Lexicon, which classifies words into eight primary emotions: anger, fear, anticipation, trust, surprise, sadness, joy, and disgust, along with two sentiments, positive and negative [Mohammad and Turney 2013]. Each comment was assigned an emotion intensity score ranging from 0.1 to 1.0, indicating the extent to which it expressed a particular emotion. This approach allowed us to assess the emotional composition of the conversations effectively.

### 3.4 Graph-based Conversation Analysis

To analyse the evolution of emotions in online discussions, we represented conversations as Directed Acyclic Graphs (DAGs). Each node in the graph corresponds to a user comment, or posts, and edges denoting the reply relationships between comments. In this structure, a directed edge connects the replying comment to the comment it responds to, creating a hierarchical association where the original post serves as the root node. This graph-based setup allows us to assess the influence of each comment on the overall emotional tone of the discussion.

To quantify the impact of individual comments on the overall emotional tone, we computed an influence score for each node based on several factors:

- Number of Replies: Comments that receive more replies are considered more influential.
- Distance from Root Node: Comments closer to the original post are deemed more influential due to their proximity to the root.
- PageRank: This widely used algorithm ranks nodes according to their significance within the graph.
- Emotion Intensity: The strength of the emotion expressed in the comment contributed to its influence. The emotional influence ($E_m$) of each comment on the root node is given by the following equation:

*The impact of nodes in $G = (V, E, A)$ on the root node $R$ is given by:*

$$\forall V \in G - \{R\}, E_m(R) = \sum E_m(V1, V2, V3....Vn) \tag{1}$$

where:

$$E_m(Vi) = f(Ai)$$

The calculated influence scores were then aggregated to form the emotional board of the root node. This emotional board was updated as the conversation progressed, reflecting the shifting emotional landscape of the conversation.

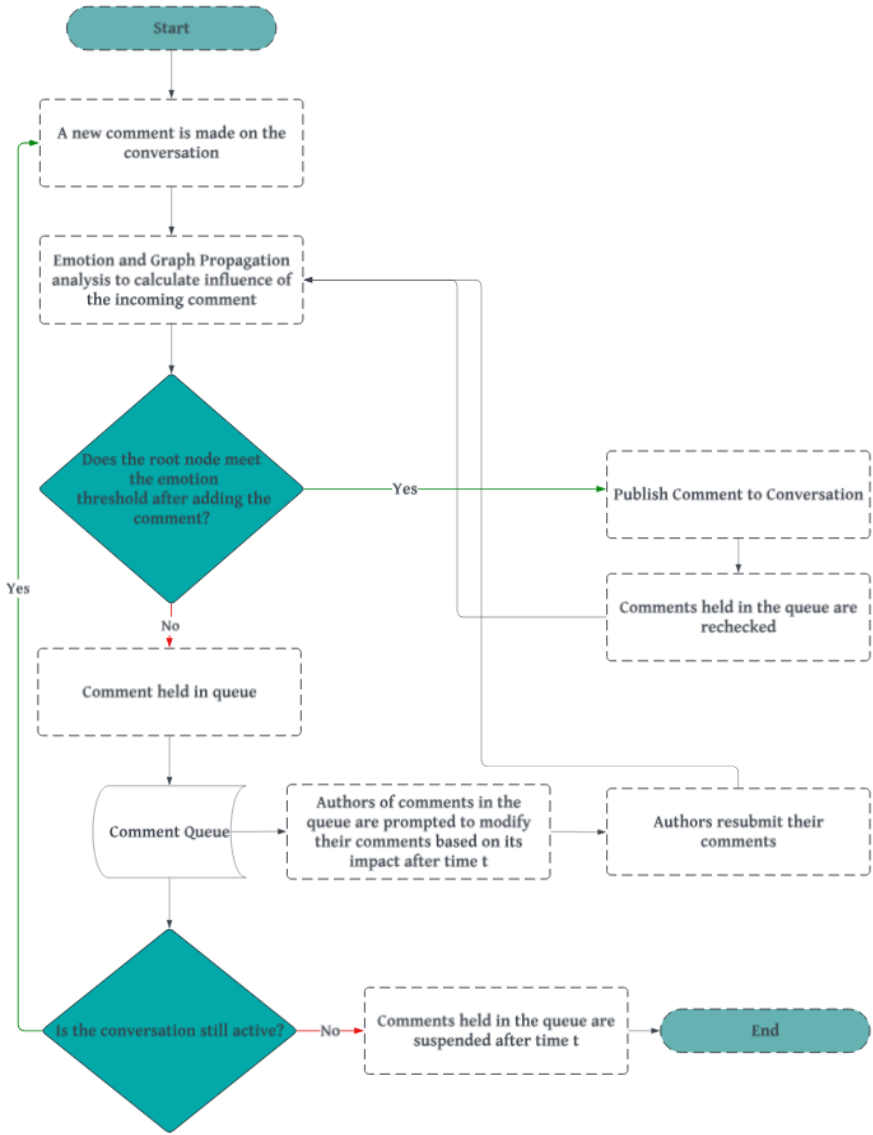

Fig. 1. Proposed Comment Queuing to encourage Self-reflection

### 3.5 Proposed Comment Queuing to encourage Self-reflection

To address the issue of negative emotions and toxic interactions within online discussions, we introduce a comment queuing mechanism designed to temporally hold certain comments prior to their publication in the conversation graph, as shown in Fig. 1. The objective of this mechanism is to facilitate self-reflection among users while simultaneously regulating the emotional trajectory of ongoing discussions.

Upon submission of a comment, the proposed framework evaluates its potential emotional impact on the conversation's emotion board. The emotion board aggregates the emotional influence of all comments in the discussion, with specific thresholds set for each emotion (e.g., Anger > 50%, Fear

> 60%). For instance, if a comment elicits an anger level exceeding 50% or a fear level surpassing 60%, it is identified as potentially disruptive. Instead of being published instantaneously, these flagged comments are placed into a queue for periodic re-assessment. They undergo re-evaluation whenever a new comment is added to the conversation.

This queuing mechanism starts with predefined thresholds for each emotion and then employs a dynamic adjustment of thresholds informed by several contextual factors, including:

- Conversation Activity Level: In instances of heightened conversational activity, thresholds may be relaxed to permit a broader emotional spectrum. Conversely, during quieter periods, thresholds may be slightly lowered to facilitate stricter moderation and prevent the potential escalation of intense emotions.
- Emotion Distribution Trends: The system observes fluctuations in emotional intensity to ensure that moderation remains responsive and adaptive.
- Sliding Window Approach: By considering only the most recent 100 comments, the system mitigates the influence of outdated emotional trends on moderation practices.

Since every comment has an impact on the overall tone of the conversation, the root node's emotion board is updated every time a new comment is added to the conversation. With each new comment submission, the system reassesses the queued comments to evaluate any changes in their emotional impact. When the conversation yields a more balanced emotional tone, previously withheld comments may be subsequently published. Conversely, if a comment continues to be emotionally disruptive, it will remain in the queue.

In instances where a comment is retained in the queue for an extended duration, the system prompts the user to consider revising their submission prior to re-evaluation. Once the user returns their comment (modified or not), the comment undergoes re-evaluation, considering any changes to the emotion board and thresholds. If the revised comment's emotional impact falls within acceptable limits, it is added back to the conversation. Conversely, if it still surpasses toxicity thresholds or the user chooses not to revise, the comment will be suspended to prevent further emotional escalation in the discussion. This approach ensures effective moderation of potentially inflammatory content while fostering constructive participation.

This adaptive queuing approach serves to effectively moderate online discussions while preserving user engagement levels. By decelerating the integration of emotionally charged comments, the system curtails the phenomenon of emotional contagion, wherein highly charged comments elicit similarly extreme reactions from other participants. Additionally, by promoting a moment of pause and reflection on the part of users, the mechanism cultivates more deliberate and considered interactions, thereby reducing impulsive emotional outbursts as well as instances of both intentional and unintentional trolling.

### 3.6 Evaluation

This section describes the process of evaluating the proposed queue mechanism. To evaluate the effectiveness of the comment queuing mechanism, we performed an empirical analysis based on data collected from Reddit. Subsequently, we conducted a user study to assess the feedback of the participants on the queuing mechanism.

*3.6.1 Empirical Analysis using Reddit Data.* To assess the effectiveness of the comment queuing mechanism, we conducted an empirical analysis using the data collected from Reddit. Our focus was on measuring the duration of comment holds (the length of time comments were delayed before publication), analysing the ratio of held comments compared to those published immediately, and examining the effect on emotional balance by comparing conversations with and without the queuing mechanism. To quantify the impact of the queuing mechanism, we tracked the distribution

Queueing for Civility: User Perspectives on Regulating Emotions in Online Conversations 9of held durations and analysed changes in the root node's emotional state over time. For this experiment, the sliding window was focused on the last 100 comments to capture the latest emotional trends and used dynamic emotion thresholds. The second part of the evaluation of the proposed framework employs a survey designed to collect respondents' perceptions regarding the framework.

*3.6.2 Understanding User Perspectives.*

*Participant Recruitment.* A user survey was conducted to collect feedback on the queuing mechanism and its potential role in moderating online conversations. Participants were recruited through social media platforms and online community forums, with eligibility criteria requiring them to be at least 18 years old, daily users of social media, and residents of Australia. The study aimed to focus on Australian adults who frequently use social media, as they are both an important demographic and convenient for sampling purposes. We recruited 20 participants from various social media platforms: 14 were employed full-time or part-time, 5 were students, and 1 was unemployed. Recruitment materials highlighted the anonymous nature of the study and emphasised the importance of participant feedback in supporting healthier online interactions. Ethics approval was obtained from Deakin University (Australia), ensuring that all participant data would be handled responsibly.

*Survey Design and Procedure.* The survey was developed using Qualtrics and comprised 17 questions, utilising Likert-scale ratings, multiple-choice formats, and open-ended responses to capture both quantitative and qualitative insights. The primary areas of focus included:

- Emotional State & Online Behaviour: Participants shared their emotional experiences while using social media (Q1) and reflected on past instances of emotional distress (Q2).
- Perception of the Queuing Mechanism: Participants shared their opinions regarding the practice of delaying responses for the sake of self-reflection (Q3–Q6).
- Impact on Conversation Dynamics: Users considered how the queuing system influenced their emotional reactions and whether it fostered more thoughtful discussions (Q7–Q10).
- Effectiveness in Reducing Toxicity: The survey explored whether the queuing feature helped mitigate impulsive comments and online trolling (Q11–Q12).
- Potential Implementation: Participants expressed their interest in seeing this feature incorporated into real social media platforms (Q16) and provided suggestions for improvements (Q15, Q17).

To enhance clarity, the survey included video demonstrations showcasing online conversations both with and without the queuing mechanism. A complete list of survey questions is shown in Table 1.

The survey was conducted entirely online, enabling participants to engage at their convenience. The process included the following steps:

- Reading an introductory description of the queuing mechanism.
- Watching a brief explanatory video that demonstrated its real-time application.
- Completing the survey, which featured Likert-scale, multiple-choice, and open-ended questions.

As a token of our appreciation, we offered a $50 Coles (a popular Australian supermarket chain) voucher to each participant who completed the study.

Quantitative responses from the survey were analysed using descriptive statistics, with an emphasis on response distributions and correlation analysis. A Chi-square test was performed to investigate the relationships between emotional state and perception of the queuing mechanism as well as the likelihood of engaging in impulsive behaviour and the perceived usefulness of the

, Vol. 1, No. 1, Article . Publication date: May 2025.

Table 1. Survey Questions and Response Types

| Survey Question | Response Type |
| --- | --- |
| How would you describe your typical emotional state when using social media? | Multiple Choice |
| Have you ever experienced emotional distress in online discussions? | Yes/No + Open-ended |
| How did you feel about the comment queuing mechanism that delays responses for self-reflection? | Likert Scale |
| Did the delay in posting comments help regulate emotions? | Likert Scale |
| Did you find the delay helpful or frustrating? | Likert Scale |
| Do you think the queuing mechanism would reduce impulsive comments? | Likert Scale |
| Do you think the queuing mechanism would help maintain a balanced emotional tone in discussions? | Likert Scale |
| How did the delay affect your willingness to engage in the conversation? | Likert Scale |
| Did the queuing mechanism encourage more thoughtful responses? | Likert Scale |
| Did you feel that the queuing system helped mitigate online trolling? | Likert Scale |
| Do you think the queuing system would be effective in reducing online toxicity? | Likert Scale |
| Would you like to see this feature implemented in real social media platforms? | Likert Scale |
| Would you be more likely to modify your comment if informed about its emotional impact? | Likert Scale |
| What improvements or changes would you suggest for the queuing mechanism? | Open-ended |
| Any additional comments or thoughts? | Open-ended |

queue. Qualitative responses (Q17) were examined thematically to highlight user concerns and suggestions for improvement.

## 4 Results and Discussion

The results and discussion section provides an analysis of the evaluation conducted on the proposed comment queuing mechanism, integrating findings from both computational experiments and user survey responses. This section aims to assess the effectiveness of the queuing system in moderating online discussions by delaying potentially harmful comments to promote self-reflection. The computational analysis investigates how the queuing mechanism influences emotional dynamics within online conversations by using real-world conversations from Reddit to compare the emotional distribution both with and without the queuing in place. Furthermore, the user survey offers both qualitative and quantitative insights regarding participants' perceptions of the mechanism's impact on conversation tone, emotional regulation, and trolling behaviour.

### 4.1 Comment Queuing Mechanism

We analysed the comment queuing mechanism using a dataset comprising 15,000 user posts and interactions collected from Reddit. Each conversation was preprocessed, underwent emotion classification, and was then represented as a Directed Acyclic Graphs (DAGs), as described in Section 3. Leveraging the timestamps associated with posts and comments, we examined the influence of the queuing mechanism on the emotional trajectory of online conversations by comparing two experimental conditions:

- Without the Queuing Mechanism: In this scenario, conversations proceeded uninterrupted, with the emotional impact of each incoming comment on the root node's emotion board being assessed.
- With the Queuing Mechanism: Here, each new comment underwent evaluation prior to its integration into the conversation, ensuring that it did not disrupt the conversation's emotional balance. Comments exceeding predefined emotional thresholds were temporarily held in a queue, allowing for the processing of subsequent comments. The queued comments were continuously re-evaluated every time a new comment was added to the conversation.

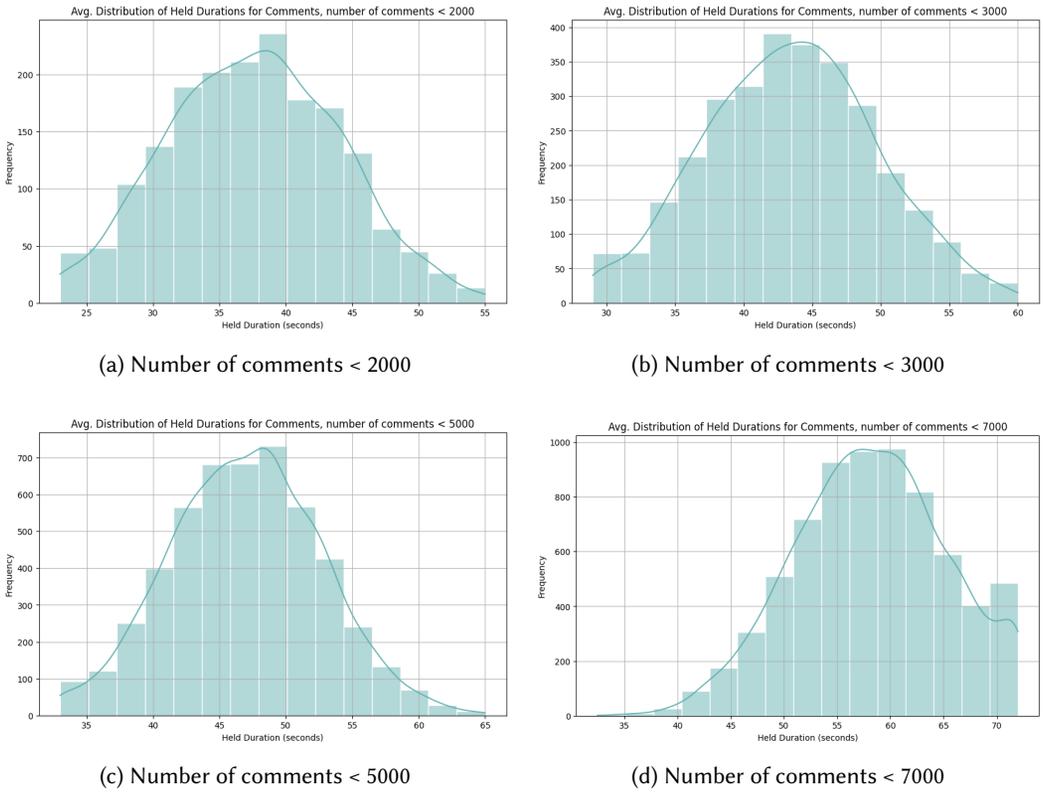

Fig. 2. Average Distribution of Held Durations for Comments when using the proposed queue approach. (a) <2000 comments, (b) <3000 comments, (c) <5000 comments, (d) <7000 comments.

The queuing mechanism was designed to dynamically adjust emotional thresholds based on multiple factors, ensuring adaptive and context-sensitive moderation:

- Conversation Size: Larger discussions permitted higher emotional thresholds, accommodating broader range of emotional variation without immediate intervention.
- Current Emotion Distribution: The prevalence of specific emotions within the discourse influenced if a new comment could be smoothly integrated.
- Engagement Level: Conversations characterised by high engagement were recognised as more volatile, thereby exhibiting increased emotional variability.
- Sliding Window Approach: The system prioritised the 100 most recent comments, thereby ensuring that moderation efforts remained relevant to the dynamic nature of the conversation while preserving the flow of conversation.

When there are no new comments to be added and some previous ones stay in the queue, the system gradually adapts the thresholds to reintegrate them. It takes into account the emotional impact of the comments and the engagement levels. In a real-time scenario, if comments remained queued indefinitely, users would be prompted to edit their response or notified of its suspension.

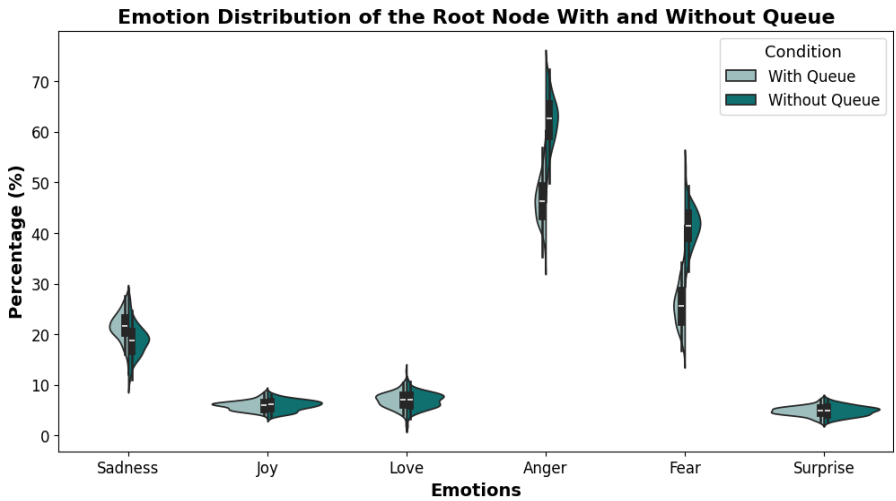

Fig. 3. The violin plot presents the distribution of emotions in the emotion board of the root node for online conversations with and without the queuing mechanism.

To analyse the delays introduced by the queuing mechanism, we plotted a histogram of held durations (Fig. 2), illustrating how long comments were retained before being posted. The X-axis represents the duration for which a comment was held (in seconds) and the Y-axis shows the frequency of comments held for that duration.

For this analysis, we categorised conversations by size and observed that most comments were held for relatively short durations, averaging between 40 to 55 seconds, which minimised disruption to the conversation. However, a small proportion of comments experienced longer delays, suggesting that they had a greater emotional impact and necessitated more contextual balance prior to their reintegration. These findings are consistent with previous research on online toxicity, indicating that a significant amount of toxicity arises from a small fraction of comments [Saveski et al. 2021], [Guberman et al. 2016]. Furthermore, the analysis showed that as the size of the conversation increased, the number of temporarily held comments also increased, from 200 in conversations containing fewer than 2,000 comments to 1,000 in those with fewer than 7,000 comments. Despite the increase in held comments, the average hold duration went up marginally, remaining less than 20 seconds, reinforcing the notion that most comments were only mildly toxic, while a smaller subset displayed potent toxicity. The system ensured a smooth conversation flow, moderating only 4% of comments, with most delays lasting less than a minute.

To assess the impact of the queuing mechanism on emotional dynamics, we conducted a comparison of emotion boards with and without the queue as shown in the violin plot(Fig. 3). The X-axis indicates the different emotions (Anger, Fear, Joy, Love, Surprise, Sadness) and the Y-axis shows the percentage contribution of each emotion to the conversation. The wider sections of each violin indicate a higher concentration of comments with that emotional intensity, while the narrower sections represent fewer occurrences. We found that in the absence of the queue, most conversations were predominantly characterised by negative emotions, particularly anger and fear, which intensified over time. With the implementation of the queuing mechanism, there was a considerable increase in the prevalence of joy and love, while anger and fear experienced a significant decrease. The queue mechanism achieved an average reduction of approximately 15% in anger, not by eliminating angry comments, but by moderating their impact on the conversation's emotional tone. By ensuring that no single emotion dominated the conversation, the queuing system effectively prevented emotional contagion and the formation of echo chambers, where the escalation of negativity could remain uncontrolled. Hence, this mechanism regulated emotional trajectories, fostering constructive and balanced discussions.

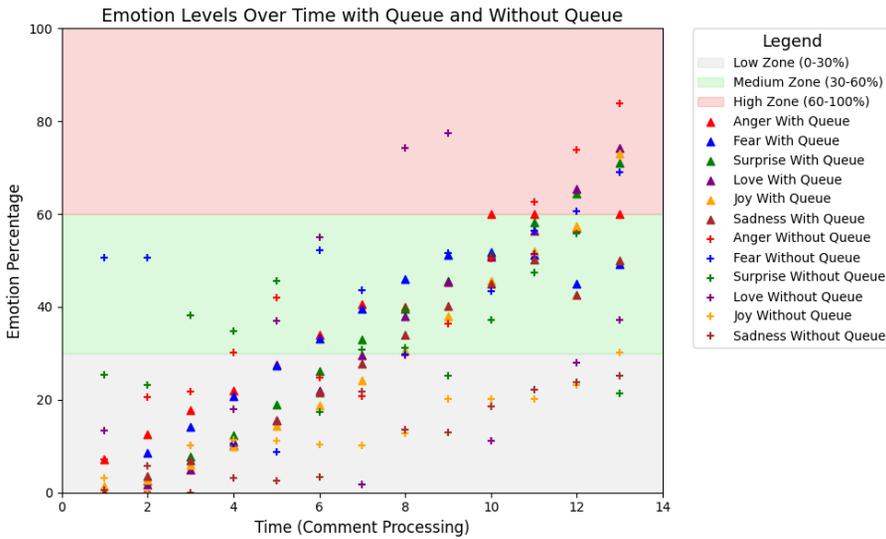

Fig. 4. Average emotion levels in the conversation when using the queue

To further explore how emotions evolved throughout the conversation, we created a scatter plot illustrating the evolution of emotions over time as shown in (Fig. 4). The X-axis represents the time (comment processing sequence) and the Y-axis shows the emotion percentage at each point in time. For ease of interpretation, we divided the plot into three distinct emotion zones. The Low (0–30%) in the grey background indicates a Stable emotional state. The Medium (30–60%) in the green background indicates Moderate fluctuations in emotions and the High (60–100%) in the red background indicates potential emotional escalation. Without a queuing system in place, anger and fear surged quickly, often crossing into high emotional territory and resulting in toxic interactions. In contrast, when the queuing mechanism was enabled, negative emotions grew at a slower pace and rarely reached the high zone, demonstrating the system's effectiveness in preventing emotional escalation. Meanwhile, joy and love showed a gradual increase, indicating that the system actively encouraged balanced emotional engagement and in turn promoted positive interactions. The sliding window and dynamic thresholds ensured the system adapted to the context of the conversation, effectively preventing abrupt and excessive moderation.

### 4.2 Survey Outcomes and Key Themes

*4.2.1 Quantitative Responses.* The survey began by asking participants to describe their typical emotional state while using social media. This was followed by an explanatory video demonstrating the proposed queuing mechanism in action. Subsequently, participants answered a series of questions about their perceptions of the delay mechanism, including whether they believed it would help regulate emotions, foster more constructive online discussions, and reduce trolling behaviour. The responses were then analysed using correlation and statistical tests to identify significant relationships between users' emotional states and their perceptions of the effectiveness of the queuing mechanism. Following were the statistical metrics we used in this study:

- Chi-Square Test: The chi-square test is a non-parametric test used to determine whether there is a significant association between categorical variables [McHugh 2013]. In this study, it assesses whether the users' emotional states influence their perception of the queuing delay. A low p-value ($p < 0.05$) indicates that the relationship is statistically significant.
- Cramér's V: Cramér's V is a measure of the effect size of categorical variables in a chi-square test [Cramér 1999]. Quantifies the strength of the association between two variables, with values close to 1 indicating a strong relationship, while lower values indicate weaker associations.
- ANOVA (Analysis of Variance): ANOVA is a statistical method used to compare the means of multiple groups to determine whether there are significant differences [Edwards 2005]. The p-value determines whether there is a statistically significant difference between the means of multiple groups. In this study, it was used to examine how different emotional states affected user perceptions of queuing delay.
- Cohen's f: Cohen's f is an effect size measure for ANOVA, used to determine the magnitude of differences between group means [Cohen 2013].
- Pairwise Comparisons: To further investigate differences between specific groups, we conducted pairwise tests, which compare two groups at a time [Maxwell 1980]. This method helps identify which specific emotional states contributed the most to the observed statistical differences.

Our statistical analysis reveals strong associations between user emotional states and their perceptions regarding the queuing mechanism's effectiveness.

- Effect of Emotional State on Emotion Regulation Through Delay: A one-way ANOVA was performed to assess the influence of participants' typical emotional states while engaging with social media on their beliefs regarding the queuing mechanism's capacity to facilitate emotional regulation (e.g., allowing for reconsideration of responses prior to posting). ANOVA assesses whether there are statistically significant differences between the means of multiple groups. To measure the practical significance of this finding, we computed Cohen's f, which quantifies the effect size of differences in means. A Cohen's f of 1.056 suggests a strong effect, meaning that emotional state substantially impacts whether users find the queuing mechanism beneficial. Larger values of Cohen's f indicate stronger group differences, reinforcing that user emotion plays a crucial role in shaping perceptions of online moderation strategies. The findings indicated a statistically significant effect ($p = 0.018$, Cohen's f = 1.056), suggesting that a user's emotional baseline significantly affects their perception of the delay as a means of emotional regulation. Participants categorising their emotional state as "calm" (M = 2.00, SD = 0.71) or "happy" (M (Mean) = 2.50, SD (St. Deviation) = 0.84) reported a greater perception of the delay as being helpful for emotion regulation compared to those who identified as "neutral" (M = 3.67, SD = 0.58) or "frustrated" (M = 4.00, N=2).
- Perceived Helpfulness of the Delay: A chi-square test was conducted to assess categorical associations and explore whether users' emotional states related to their perceptions of the delay as either helpful or frustrating. The findings indicated a statistically significant relationship between emotional states and the perception of queuing delays, with a p-value of 0.0029 and a strong effect size represented by Cramér's V of 0.83. This suggests the critical role that emotional state has in shaping users' attitudes towards the queuing mechanism. Figure 5 demonstrates the correlation between emotional states and user responses to the delay mechanism. All participants who described their emotional state as neutral, calm,

| Q1: How would y...g social media? | Q5: Did you find th...ful or frustrating? | | |
| --- | --- | --- | --- |
| | Neutral | Helpful | Frustrating |
| Neutral | 33.3% | 16.7% | 0.0% |
| Calm | 0.0% | 41.7% | 0.0% |
| Happy | 33.3% | 41.7% | 0.0% |
| Frustrated | 0.0% | 0.0% | 100.0% |
| Sad | 33.3% | 0.0% | 0.0% |
| Total | 100.0% | 100.0% | 100.0% |

(a) Chi-square results

| Group A | Group B | Difference (A - B) | Sample Size |
| --- | --- | --- | --- |
| Frustrating | Helpful | -16.7% | 14 |
| Frustrating | Neutral | -33.3% | 5 |
| Helpful | Frustrating | 16.7% | 14 |
| Helpful | Neutral | -16.7% | 15 |
| Neutral | Frustrating | 33.3% | 5 |
| Neutral | Helpful | 16.7% | 15 |

(b) Pairwise statistics

Fig. 5. Perceived helpfulness of the delay. (a) Chi-square results. (b) Pairwise statistics.

or happy found the queuing mechanism beneficial for moderating discussions. Notably, 41.7% of participants who identified as happy and 41.7% who felt calm found the delay to be advantageous. Conversely, individuals who reported feelings of frustration uniformly perceived the delay as detrimental, suggesting that those predisposed to frustration are more likely to regard the queuing system as an obstacle to engagement rather than a chance for reflection. Additionally, participants who characterised their emotional states as neutral exhibited a range of perceptions: 33.3% viewed the delay as helpful, while an equal proportion found it to be neutral. Respondents who identified as feeling "sad" while engaging with social media displayed varied responses, with 33.3% perceiving the delay positively and another 33.3% remaining neutral. In order to further analyse the relationship between emotional states and perceptions of the queuing mechanism, pairwise comparisons were performed, where Group A represents one emotional state (e.g., Frustrated, Calm, Happy, or Neutral) and Group B Represents a contrasting emotional state for comparison. The differences calculated (A - B) reveal how users from diverse emotional backgrounds responded to the queuing mechanism differently. There is a 100% difference in perception, with all frustrated participants deeming the delay frustrating, while all those who found the mechanism helpful viewed it as beneficial. A consistent 100% difference emerged, where frustrated users perceived the queue as a barrier, whereas neutral users presented a divided opinion. A 16.7% difference was observed; neutral users exhibited a greater openness to the queuing mechanism but were less likely to find it useful compared to their happy or calm counterparts.
- Impact on Conversation Tone: Another chi-square test was employed to investigate the relationship between emotional state and perceptions of the queuing mechanism's impact on conversation tone. The results were statistically significant (p=0.00193, Cramér's V = 1.0), indicating a strong relationship between emotional state and users' perceptions of whether the

| Q1: How would y...g social media? | Q13: How did you feel the delay in adding comments impact the overall tone of the conversation? | |
|---|---|---|
| | Calm the conversation | Made the conversation more frustrating |
| Neutral | 21.4% | 0.0% |
| Calm | 35.7% | 0.0% |
| Happy | 42.9% | 0.0% |
| Frustrated | 0.0% | 66.7% |
| Sad | 0.0% | 33.3% |
| Total | 100.0% | 100.0% |

(a) Chi-square results

| Group A | Group B | Difference (A - B) | Sample Size |
|---|---|---|---|
| Calm the conversation | Made the conv... | 21.4% | 17 |
| Made the conversatio... | Calm the conv... | -21.4% | 17 |

(b) Pairwise statistics

Fig. 6. Effect of Emotional State on users' perception of how the delay in adding comments impacts the overall tone of the conversation. (a) Chi-square results. (b) Pairwise statistics.

queuing mechanism improved discussion quality. Fig. 6 illustrates the relationship between emotional state and perceived conversation tone. Among participants who described their typical emotional state as neutral, calm, or happy, 100% believed that the queuing mechanism contributed to a calmer conversation. In contrast, the only participants who perceived the queuing mechanism as making the conversation more frustrating were those who identified as feeling frustrated or sad. Notably, 66.7% of frustrated users felt that the queuing system exacerbated the conversation's frustration, while 33.3% of sad users shared this perspective. These findings suggest that while the queuing mechanism can help regulate discussion tone for most users, individuals prone to frustration or sadness may interpret the delay as an additional barrier rather than an opportunity for self-regulation. To further investigate differences between specific emotional states, ranked pairwise comparisons were conducted where Group A represents users who felt the queuing mechanism calmed the conversation and Group B represents users who felt the queuing mechanism made the conversation more frustrating. The difference (A - B) measures how different emotional states contributed to users' perceptions of the queuing system. All frustrated users perceived the queuing mechanism as frustrating, while all calm users found it beneficial (66.7% difference). While most happy users found the mechanism calming, their responses showed slightly more variation than those of neutral or calm users (42.9% difference). Neutral users were generally inclined to view the queue positively, but some remained uncertain about its effectiveness (21.4% difference). While some sad users acknowledged the queue's role in moderating emotions, a third of them still found it frustrating (33.3% difference).

In general, a substantial majority (93.3% participants) felt that the proposed mechanism would mitigate trolling and impulsive comments, expressing a desire for its implementation on social media platforms. Additionally, 83% of participants believed that the delay mechanism would contribute to a more balanced emotional tone in online discussions by deterring impulsive, emotionally charged responses.

*4.2.2 Qualitative Responses.* In the survey, participants were asked to suggest improvements or changes they believe would improve the queuing mechanism. They provided a range of feedback regarding potential enhancements to the queuing mechanism aimed at better regulating emotions in online conversations. The key themes gathered from their responses can be summarised as follows.

- Transparency and User Awareness: A number of respondents felt that increasing transparency by providing explicit rationales for why a comment is queued could significantly enhance user understanding and foster self-reflection. For instance, a brief notification indicating the reason for a comment being held (e.g., "Your comment is queued due to emotional tone") could improve clarity regarding the queuing process. Furthermore, the introduction of a visual indicator, such as a mood gauge, was recommended to track the emotional tone of conversations in real-time, thereby enabling users to observe emotional trends and adjust their responses accordingly.
- Enhancing Emotional Self-Regulation: Participants proposed the implementation of cool-down periods wherein emotionally charged messages would be temporarily delayed (e.g., for 30 seconds) to afford users the opportunity to reconsider and amend impulsive responses. Additionally, participants suggested implementing empathy prompts, such as 'How might this message affect the recipient?', to encourage users to reflect on their tone before posting. The use of AI-powered interventions, particularly sentiment analysis tools, was also recommended to detect emotional intensity, providing users with suggestions for alternative phrasing or calming resources.
- Personalised and Adaptive Mechanisms: Several respondents supported the use of an adaptive queuing system that takes into account individual emotional states, recommending breaks or redirection when users appear visibly upset. Moreover, a priority system wherein neutral or positive messages are prioritised over highly emotional or aggressive comments was proposed to promote constructive discourse. Gamification elements, such as badges, points, or titles (e.g., "Empathy Leader"), were suggested to reward respectful engagement and reinforce positive behavioural patterns.

Overall, while participants acknowledged the potential benefits of the queue in mitigating impulsive comments and reducing toxicity, they highlighted the need for greater transparency and adaptability to ensure a seamless user experience.

These findings reinforce the effectiveness of the queuing mechanism in promoting healthier online interactions. They validate the contributions of the paper by providing empirical evidence supporting the efficacy of the proposed queuing mechanism. The findings confirm that delayed responses can regulate emotions, improve digital civility, and maintain a balanced conversation through dynamic adaptation. These insights add to the existing literature on emotion-aware content moderation and create avenues for real-time intervention strategies within online platforms.

## 5 Limitations

This study demonstrates that a comment queuing system can help reduce impulsive and harmful interactions in online discussions. However, there are some important limitations to keep in mind.

First, the evaluation used data from Reddit, which means the queuing system was tested in a simulated setting instead of a real one. While the results suggest that this system can reduce negative feelings and improve emotional balance, implementing it in real life could be more complicated. Users might find ways to bypass the system, and engagement levels may vary across different platforms. The study also focuses exclusively on English-language content, which may limit the generalisation of the findings.

Second, the user survey included a small group (N=20) of people who were already interested in regulating online discussions. Although their views are useful, a larger and more diverse group would provide stronger insights. Additionally, the study mainly focused on comments in English and may not account for the complexities of multilingual or culturally diverse conversations.

Third, while the study suggests that delays in posting comments may encourage users to think before they post, the long-term effects of this behaviour are still unclear. It is uncertain whether users will change their commenting habits over time or if repeated use of the queuing system might lead to frustration or disengagement. Long-term studies and real-world tests will be essential to understand how effective the system is over time and in different online communities.

In conclusion, despite these limitations, this study takes an important step toward creating real-time tools for managing emotions in online interactions. It highlights the need for ongoing research to find the right balance between moderation, free speech, and user experience.

## 6 Conclusion and Future Work

This study introduced an innovative comment queuing mechanism designed to regulate emotional intensity in online discussions and mitigate the effects of impulsive and toxic interactions. By analysing 15,000 instances of Reddit conversations, we demonstrated that implementing a queue-based delay can significantly reduce the spread of negative emotions, such as anger and fear, while preserving a balanced emotional tone. The system's adaptive thresholds and sliding window approach effectively managed real-time emotional fluctuations, preventing emotionally charged comments from dominating discussions.

In addition to the computational evaluation, we conducted a user survey to gather insights into how the queuing mechanism is perceived in real-world online interactions. The findings showed support for the system, with 93% of participants agreeing that the delay would help to calm conversations and 83% believing it could reduce impulsive comments. Furthermore, a statistically significant relationship was identified between users' emotional states and their perception of the queuing mechanism, stressing the importance of tailored interventions. Qualitative feedback also suggested enhancements such as real-time feedback, emotional check-ins, and AI-assisted moderation to improve system usability.

Overall, the results indicate that delaying emotionally charged comments encourages self-reflection, promotes more constructive engagement, and reduces the likelihood of escalation in online discussions. Future work will focus on refining the queuing mechanism, incorporating context-aware interventions, and testing it within live online communities to assess its long-term impact on digital discourse.

During the drafting of this paper, [Grammarly 2024] was used to check and enhance the grammar and writing style of this document.